\documentclass[noshowpacs,prl,aps,superscriptaddress,floatfix, twocolumn,footinbib]{revtex4-2} 
\usepackage[utf8]{inputenc}
\usepackage{lineno,mathtools,url}
\usepackage{amsmath}
\usepackage{soul}
\usepackage{amssymb,mathrsfs}
\usepackage{physics}

\usepackage{graphicx}    \usepackage{cuted} 
\usepackage{hyperref}

\usepackage{comment}
\usepackage{dsfont}
\usepackage[dvipsnames]{xcolor}
\hypersetup{
     colorlinks=true,
     linkcolor=blue,
     filecolor=blue,
     citecolor = orange,      
   }

\newcommand{\px}{\partial_x}
\newcommand{\pt}{\partial_t}
\newcommand{\rhon}{\rho'_n}
\newcommand{\rhou}{\rho_n^u}

\newcommand{\vcr}{v_n^{\text{cr}}}
\usepackage{mathbbol} 
\usepackage[normalem]{ulem}

\allowdisplaybreaks
\begin{document}

\title{
Anomalous Doppler effect in superfluid and supersolid atomic gases
}

\author{Tomasz Zawi\'slak}
\author{Marija \v{S}indik}
\author{Sandro Stringari}
\author{Alessio Recati}\email[]{Corresponding Author: alessio.recati@ino.cnr.it}
\affiliation{Pitaevskii BEC Center, CNR-INO and Dipartimento di Fisica, Universit\`a di Trento, Via Sommarive 14, 38123 Povo, Trento, Italy}

\begin{abstract}

By employing the formalism of hydrodynamics, we  derive novel analytic predictions for the 
Doppler effect  in superfluids  with broken Galilean invariance and hosting  persistent currents at zero temperature. 
We consider two scenarios: when Galilean invariance is broken explicitly (by external potentials) and
spontaneously, as it happens in a supersolid. In the former case, the presence of a stationary current affects
the propagation of sound via an anomalous Doppler term proportional to the density
derivative of the superfluid fraction. In supersolids, where, according to Goldstone theorem, distinct
sounds of hybrid superfluid and crystal nature can propagate, the Doppler effect can be very
different for each sound.
Quantitative estimates of the Doppler shifts are obtained for Bose-Einstein condensed atomic gases, described by Gross-Pitaevskii theory.  The estimates are obtained both calculating  the thermodynamic parameters entering the hydrodynamic results, and from full time-dependent simulations.

\end{abstract}

\maketitle

In a classical fluid moving at a small velocity $v_f$, the sound speed $c_0$ is modified by the kinematic Doppler shift as $c=c_0\pm v_f$, 
depending on whether the sound propagates parallel ($+$) or antiparallel ($-$) to the velocity $v_f$. The Doppler effect has been predicted to be different in a superfluid due to the relative velocity between the normal and the superfluid component. 

This problem was first addressed a long time ago by Khalatnikov~\cite{Khalatnikov56} 
for superfluid helium at finite temperature, where the motion of the liquid is described by Landau's two-fluid hydrodynamics. Stimulated by early
experimental results on the Doppler effect in the propagation of fourth sound in Helium~\cite{Rudnick1969}, further theoretical studies have pointed out  
the occurrence of non-trivial, anomalous Doppler shifts in $^4$He~\cite{Bergman1975,NepoHeII, NepoHe4thSound},  
$^4$He$-^3$He mixture~\cite{NepoHe34},  as well as in superfluid $^3$He~\cite{kenis99}.
To our knowledge, however,  a clear experimental confirmation of the anomalous  Doppler effect in  
liquid helium is still missing. 

In this Letter, we demonstrate the possibility of a zero-temperature Doppler effect  in ultra-cold gas superfluids~\cite{Varenna1999,RMPBloch2008,BecBook2016}, 
which, differently from liquid Helium,  allow for a microscopic treatment based on Gross-Pitaevskii theory. We focus on density-modulated  configurations, where  the superfluid density can differ significantly from the average density 
even at zero temperature. Density modulated phases in ultra-cold atomic gases can be the result of an external periodic potential 
(optical lattices~\cite{Lewenstein2016}) or of the spontaneous breaking of translational symmetry, yielding  supersolidity \cite{Recati2023}. 
The propagation of sound  and the occurrence of collective oscillations,  in the absence of steady currents,  
have already been the subject of experimental investigations in both cases (see~\cite{Fort2003,Chauveau2023,Tao2023} and~\cite{F2,I2,S2}, respectively). 
Furthermore, the Doppler effect was observed in a uniform Bose-Einstein condensed gas confined in a ring,  where it was used to measure the quanta of circulation 
characterizing the velocity of the superfluid flow~\cite{Kumar2016}.

In the presence of an external periodic potential, the normal component of the fluid is locked, and only the superfluid component  can move. 
Instead, in supersolid configurations,  both superfluid and normal (crystal) components can move. As predicted in the seminal paper by 
Andreev and Lifshtz~\cite{AndreevLifshitz}, Goldstone modes, of hybridized crystalline and superfluid nature, are expected to occur in a supersolid 
(see  also~\cite{Brauner2012} ). 
 These modes are expected to react differently to the presence of a relative motion 
between the two components.

We provide general analytical results for Doppler shifts at zero temperature within a hydrodynamic approach, which, in the case of a supersolid, have never been obtained before. Considering ultra-cold gases platforms, we use  Gross-Pitaevskii theory to confirm the validity of the hydrodynamic predictions and to determine the parameters entering the hydrodynamic equations. In particular, concerning supersolids,  we consider the dipolar gas platform~\cite{ETH_SS1,Li2017,F1,I1,S1}  (see also~\cite{ Recati2023reviewSS} and references therein).

In the following, for the sake of simplicity, we consider a linear tube geometry of length $L$, with periodic boundary conditions, where we can use
one-dimensional hydrodynamic theories.
In cold atomic gases, quasi-one-dimensional configurations hosting  permanent currents can be experimentally realized using 
ring-shaped trapping potentials 
with sufficiently large radii.

\paragraph{Doppler shift of fourth sound.}

One peculiar sound that propagates in superfluid   helium occurs when the normal component is at rest due to viscous drag in porous media \cite{Pellam} or in a narrow channel \cite{Atkins}. This mode is usually referred to as fourth sound. An analog mode -- hereafter also called fourth sound -- propagates at low temperature in a Bose-Einstein condensed atomic gas in the presence of a periodic external potential, which causes the pinning of the normal component. 

Doppler effect in fourth sound was observed in helium in a superleak~\cite{Rudnick1969,Kojima1971}. In these pioneering experimental works, it was suggested that the Doppler shift has a kinematic nature and is fixed by the fluid velocity $v_f=v_s\rho_s/\rho$,  given by the current divided by the mass density $\rho$, with $\rho_s$ and $v_s$ the superfluid density and the superfluid velocity, respectively. However, a more detailed analysis based on two-fluid hydrodynamics showed that this prediction is in general not correct~\cite{NepoHe4thSound}. 
The correct result for the Doppler effect of fourth sound at zero temperature can be derived within the Hamiltonian formalism for two fluid hydrodynamics \cite{Pokrovski1976}. Assuming that the stationary superfluid velocity is small, the energy density of the system, in the  frame where the normal component is at rest, is given by the expansion  $\epsilon(\rho,v_s) =\epsilon_0(\rho) + (1/2)\rho_sv^2_s$, where $\epsilon_0$ is the energy at rest.  The superfluid velocity is irrotational and can be written as $v_s=\hbar \px \phi/m$, where the macroscopic phase $\phi$ is canonically conjugated to the density \footnote{While for Bose superfluids the mass $m$ coincides with the atomic mass, for a Fermi superfluid $m$ is twice the atomic mass.}. The Hamiltonian equations of motion for $\rho$ and $v_s$ then take the form of the following (collisionless) hydrodynamic (HD) equations:
\begin{eqnarray}\label{eq:hdol}
    \partial_t \rho+\px (\rho_sv_s)&=&0\\
    \partial_t v_s +\px\left( {\mu_0} +\frac{1}{2} \frac{\partial\rho_s}{\partial  \rho} v_s^2\right)&=&0
\end{eqnarray}
where $\mu_0(\rho) = \partial \epsilon_0/\partial \rho$ is the chemical potential calculated at $v_s=0$. The two HD equations correspond to the atom number conservation and to the presence of superfluid long-range order.

Sound propagation -- corresponding to the superfluid Goldstone mode -- is described by linearizing the above equations around the uniform equilibrium values $\rho_0$, $\rho_s^0$ and  $v_s^0$, as $\rho(x,t) = \rho_0+ \delta \rho(x,t)$, $\rho_s(x,t)=\rho_s^0+(\partial \rho_s/\partial \rho) \delta  \rho(x,t)$ and $v_s(x,t)=v_s^0+\delta v_s(x,t)$. In this way we find that the excitations obey the phononic dispersion relation $\omega(q)=c|q|$,  with $q$ the wave vector of the sound wave and $c$  the speed of sound. The latter is  given  by $c^{\pm}=c_0\pm\Delta c$, where $c_0=\sqrt{f_s\kappa^{-1}}$ is the zero-temperature value of the fourth sound velocity in the absence of current, with $f_s=\rho_s/\rho$ the superfluid fraction and $\kappa^{-1}= \rho\partial_\rho \mu_0$ the inverse compressibility \footnote{An experimental verification of the hydrodynamic relation between $c_0$ and $f_s$ was obtained only recently in~\cite{Chauveau2023,Tao2023}.}. Finally, the Doppler shift reads
\begin{equation}
\Delta c= \frac{\partial \rho_s}{\partial \rho}v_s^0=\left(f_s+ \rho\frac{\partial f_s}{\partial \rho}\right)v_s^0.
\label{eq:Doppler_ol}
\end{equation}
Therefore, due to the density dependence of the superfluid fraction, the Doppler shift differs in general from the kinematic expectation $f_s v_s^0$ .
The expressions given by Eqs.~(\ref{eq:hdol}-\ref{eq:Doppler_ol}) hold at zero temperature for  both Bose and Fermi superfluids, provided the superfluid velocity is small enough, namely $v_s^0 \ll c_0$. Similar results were obtained in~\cite{Bergman1975} in the case of superleak helium configurations, with special focus on finite size effects,  and in~\cite{smerzitrombet2003,Taylor2003} in the case of atomic Bose-Einstein condensates in optical lattices, after  linearization of  the time-dependent Gross-Pitaevskii equation (GPE). The superfluid fraction in these latter works has to be interpreted in terms of effective tunneling or effective mass.

The result of Eq.~(\ref{eq:Doppler_ol}) can be  easily generalized to the case of  an optical lattice moving with velocity $v_n^0$, by a simple Galilean transformation, i.e., $\Delta c= {v_n^0+(\partial \rho_s/\partial \rho) (v_s^0-v_n^0)}={(\partial \rho_s/\partial \rho) v_s^0+ (\partial \rho_n /\partial \rho) v_n^0}$, with $\rho_n=\rho-\rho_s$ the normal density, thereby revealing that anomalous Doppler shifts take place only if the two fluids move with different velocities.

\begin{figure}[!ht]
    \centering
    \includegraphics[width=\linewidth]{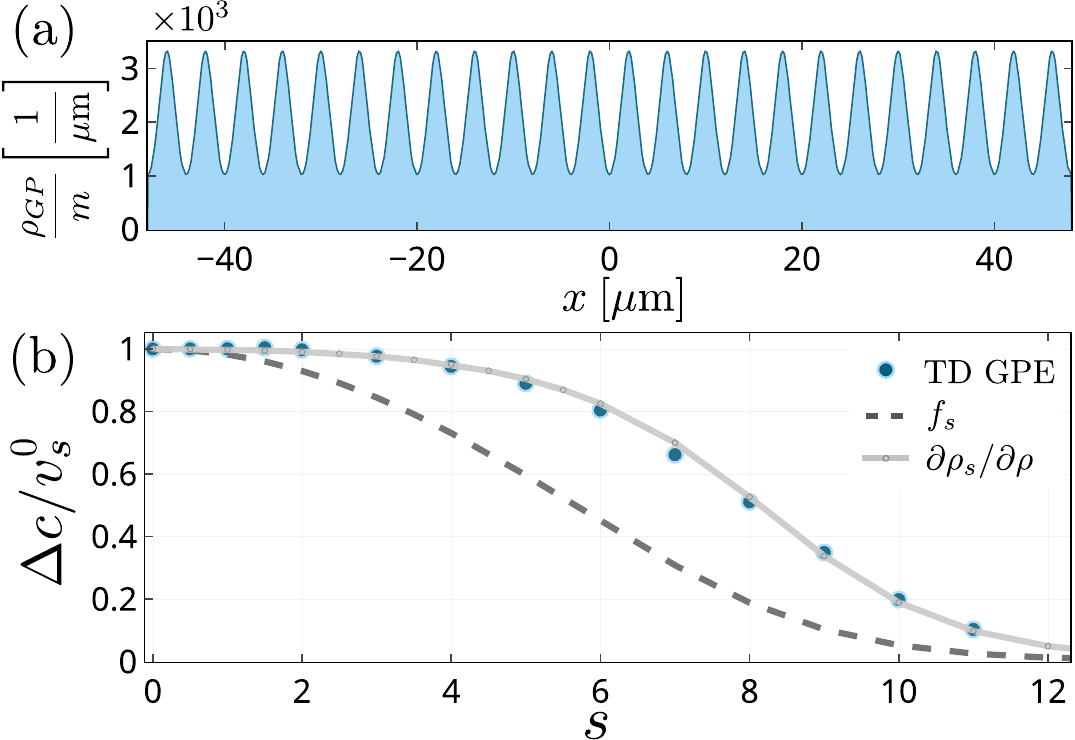}
    \caption{ (a) Example of a transversely integrated fluid density in an optical lattice of strength $s=3$.  
    (b) Doppler shift relative to $v_s^0$ of  fourth sound in an optical lattice as a function of the strength parameter $s$. Numerical results of time-dependent GPE (points) are compared with the hydrodynamic prediction from Eq.~(\ref{eq:Doppler_ol}) (solid line) and  with the kinematic estimate ${\Delta c/v_s^0=f_s}$ (dashed line). }
    \label{fig:opticallattice}
\end{figure}
We verify the validity of  the hydrodynamic result Eq.~(\ref{eq:Doppler_ol}), when applied to a Bose-Einstein condensed gas, by numerically solving the time-dependent GPE in the presence of an optical lattice. The two approaches are expected to be consistent in the phonon long wavelength limit. 
We consider realistic parameters for a gas  of $N=2\cdot 10^5$ $^{87}$Rb atoms confined in a  tube of length $L=96\mu \text{m}$ along the $x$-axis, subject to periodic boundary conditions, and radially confined by a harmonic potential of the form  $1/2 m\omega_{\perp}^2r^2$, with $r^2=y^2+z^2$, $\omega_{\perp} = 2\pi \cdot 150$Hz, with $\sqrt{\hbar/m\omega_\perp}\ll L$. The gas is subjected to an external periodic  potential along $x$ of the form   $V_{ext}=sE_R \cos (qx)$,  where $E_R=\hbar^2 q^2/2m$ is the recoil energy, and the dimensionless parameter $s$ gives the strength of the external potential. 
At zero temperature, the gas forms a Bose-Einstein condensate, characterized by a healing length $\xi=\hbar/(\sqrt{2 m\mu_0 })$, which represents the length scale on which the condensate can react to an external potential.  We choose the period of the lattice $d=2\pi/q=4\mu$m to be much larger than the healing length: $d \approx 16.7\xi$. In the opposite regime, $d\ll \xi$, the superfluid fraction becomes density independent and the kinematic Doppler shift is recovered~\cite{smerzibishop2002}. An example of the transversely integrated GPE ground state  density ${\rho_{GP}} (x)$ is reported in Fig.~\ref{fig:opticallattice}(a).
A permanent current,  providing a superfluid velocity $v_s^0=47.8\mu\text{m/s}$, satisfying the proper boundary condition, is applied to the  system.  Following a previously proposed method for probing excitations \cite{Kumar2016,Sindik2023SoundSA}, the system is prepared in a stationary state in presence of a small additional potential  $\lambda \cos (k x)$, which is suddenly removed. Since we are interested in the lowest energy modes we take the smallest possible value $k=2\pi/L$, corresponding to a wavelength much larger than the transverse size of the gas.  Within linear response theory, the time evolution of the average $\langle \cos (k x)\rangle$ contains the frequencies of the modes excited by the perturbation.  In the absence of current, the signal contains a single frequency $\omega$, from which the speed of sound is determined as $c=\omega/k$. The current induces a mode splitting ($\omega^{+}\neq \omega^{-}$), from which the Doppler shift $2\Delta c=(\omega^+-\omega^-)/k$ can be extracted.

In Fig.~\ref{fig:opticallattice}(b), we report the Doppler shift of fourth sound predicted by time dependent GPE (dots), along with the HD prediction Eq.~(\ref{eq:Doppler_ol}) (solid line), as well as with the  kinematic estimate $\Delta c/v_s^0=f_s$ (dashed line). The quantities entering the HD equations are coarse-grained and correspond to averages over the distances characterizing the microscopic density modulations of the system. A typical example of a coarse-grained average is the Leggett's  upper bound for the superfluid density~\cite{Leggett1970,Leggett1998}: $L\rho_s^{-1}=\int dx/\rho_{GP}(x)$. It provides a very accurate estimate for $\rho_s$ in the present configuration (see, e.g.,~\cite{Smith2023}).
The quantity $\partial \rho_s/\partial \rho$ entering the Eq.~(\ref{eq:Doppler_ol}) is obtained by determining the ground state of the system and extracting the superfluid density using Leggett's upper bound for different atom numbers. For the chosen parameters we see that there exists a broad region in the lattice depth $s$ where the kinematic shift substantially underestimates the Doppler effect.

\paragraph{Doppler effect in a supersolid.}  

Supersolids are systems which present both superfluid and crystalline long-range order, which, consequently, show two classes of phononic (Nambu-Goldstone) modes~\cite{Brauner2012}, due to the simultaneous and spontaneous breaking of $U(1)$ symmetry and Galilean invariance herafter called first ($c_1$) and second ($c_2$) sounds.
With respect to the case of fourth sound, the HD theory of supersolid includes additional equations for the current (following from momentum conservation) and for $u_x \coloneqq \px u$ , where $u$ is the lattice displacement (following from
crystalline long-range order) (see, e.g.,~\cite{Son, Liu_PhysRevB.18.1165, Saslov_PhysRevB.15.173,Dorsey2010} and \cite{SupplementaryMaterial}): 

\begin{eqnarray}
    \pt \rho+\px j&=&0\label{eq:yd_n}\\
    \pt j+\px\left(p+\rho_n v_n^2 +\rho_s v_s^2\right)&=&0\label{eq:yd_j}\\
    \pt v_s + \px (v_n v_s+\mu) &=&0\label{eq:yd_vs}\\
    \pt u_x + \px (v_n u_x - v_n)  &=&0,\label{eq:yd_u}
\end{eqnarray}
where we remind that $\rho_n$, $v_n$ ($\rho_s$, $v_s$) are the normal (superfluid) density and velocity in the two fluid model, $\rho=\rho_n+\rho_s$ is the total density, and $j=\rho_n v_n +\rho_s v_s$ is the total current. The quantities $p$ and  $\mu$ are the pressure and the chemical potential in the presence of stationary flow \footnote{Here we wrote the chemical potential $\mu$ in the laboratory frame. Using $\mu_s - \mu = v_nv_s - \frac{1}{2}v_s^2$ \cite{PuttermanBook,KhalatnikovBook} one can recover the standard form of Eq.~(\ref{eq:yd_vs}) (see also \cite{SupplementaryMaterial})}.
Since we are interested in the linear expansion of the HD equations, the constitutive relations can be written as~\cite{Liu_2023,Dorsey2010}:
\begin{eqnarray}
    \delta\mu&=&{1\over {\rho\kappa}} \delta \rho+\gamma u_x + {\rho_s'\over 2}\delta w^2 -{1\over 2}\delta v_n^2, \label{eq:dmu}\\
    \delta p&=& \left({1\over \kappa}-\gamma\right)\delta \rho + \left(\gamma \rho -\lambda\right) u_x + \frac{\rho_n - \rho\rho_n'}{2}\delta w^2, \label{eq:dp}\\
    \delta \rho_n &=& \rho_n' \delta \rho + \rho_n^u u_x , \label{eq:drhon}
\end{eqnarray}
where we introduce the relative velocity $w\equiv v_n-v_s$, $\gamma$ the strain-density constant and $\lambda$ the elastic constant. For the sake of notation, we define ${\rho_i'\coloneqq \partial \rho_i/\partial \rho}$, with ${i=n,s}$,  and $\rhou\coloneqq \partial \rho_n/\partial u_x$. The dependence of $\rho_n$ on the velocity difference $w$ is quadratic and can be consequently neglected at the linear order. In order to calculate the speeds of sound, we linearize Eqs.~(\ref{eq:yd_n}-\ref{eq:yd_u}) around their stationary values $\{n^0, v_s^0, v_n^0, u_x^0 =0 \}$ and determine the dispersion relation
to the lowest order in the stationary velocities. We find four phononic modes $\omega_{1,2}^{\pm}=c_{1,2}^{\pm}|q|$, where the speeds of sound can be written as $
c_{1,2}^{\pm} = c^{0}_{1,2}\pm (v_s^0+ w^0\delta_{1,2}$). The expressions for the two speeds of sound at rest $c^{0}_{1,2}$  in the supersolid phase have been already derived in a number of previous works~\cite{Pomeau94,Dorsey2010, Hofmann2021,PlattSounds}  (see below). When $w^0=0$, i.e., the superfluid and normal components move together as a single fluid, both sounds are simply shifted by the kinematic Doppler effect, fixed by the fluid velocity $v_s^0=v_n^0$. When $w^0\neq 0$, we instead find the additional  contribution $w^0\delta_{1,2}$, with
\begin{widetext}
\begin{equation}
\label{eq:Doppler_ss}
	\delta_{1,2}= \frac{ \frac{\rho_n^u}{2\rho_n}\left[\left(c^{0}_{1,2}\right)^2 - c_{\kappa}^2+\gamma\right] +  2(c_{1,2}^0)^2   - \left(1 + f_s\right)c_{\kappa}^2 -\rho_n'\beta+ \gamma\left[1+\frac{\rhon}{\rho_n}(\rho-2\rho_n)\right]}
	{2\left(c^{0}_{1,2}\right)^2  - c_{\kappa}^2-\beta} , 
\end{equation}
\end{widetext} 
 where $c_{\kappa} = \kappa^{-1/2}$ and $\beta = \lambda/\rho_n$, and the speeds of sound at rest are ${(c_{1,2}^0)^2=(A\pm\sqrt{A^2-4 B})/2}$ with ${A=c_\kappa^2+\beta-2\gamma}$ and ${B=f_s(\beta c_\kappa-\gamma^2/f_n)}$, where $f_n = 1-f_s$ \footnote{A detailed derivation of the above result  can be found in the Supplementary Materials~\cite{SupplementaryMaterial}}.
 In the limit of an incompressible lattice, i.e., as ${\beta \to \infty}$, the Doppler shift approaches the value  $\rho_s' v_s^ 0+\rho'_n v_n^0$, which coincides with the result obtained above for fourth sound. 
Moreover, approaching the crystal phase, i.e., $\rho_n\rightarrow \rho$ ($\rho_n^u\rightarrow 0,\; \rhon\rightarrow 1$),  one finds $\delta_{1,2}\rightarrow 1$, confirming the intuitive result that in the crystal phase, the sound speed simply feels a kinematic Doppler shift, given by the normal velocity $v_n^0$.

In general, Eq.~(\ref{eq:Doppler_ss}) predicts that the Doppler effect affects differently the two sound speeds. This result corresponds to the zero-temperature analogue of the predictions obtained for the first and second sound velocities in superfluid $^4$He~\cite{Khalatnikov56, NepoHeII}.

We explore the HD prediction for the Doppler effect in the case of a Bose-Einstein condensed gas of atoms interacting with dipolar forces, where successful experimental achievements of supersolid configurations have been obtained  in the last few years (see, e.g.,~\cite{Recati2023reviewSS} and references therein).  Theoretically, the supersolid phase  in such systems is stabilized by the inclusion of the so-called Lee-Huang-Yang correction to the mean-field approach,  yielding an extended version of the GPE~\cite{Wachter2016}, which has been already employed to predict the sound velocities in the absence of persistent currents.  
Within such a formalism, the speeds of sound can be extracted 
following the protocol previously described for the fourth sound of a Bose-Einstein condensate in an optical lattice. In the absence of a persistent current,  we already successfully applied the protocol to the  supersolid phase of ultra-cold dipolar gases in~\cite{Sindik2023SoundSA}.

In particular, we consider $N=1.6\cdot 10^5$ $^{164}$Dy atoms confined, as before, in a tube of length $L=96\mu$m  with periodic boundary conditions and with a transverse potential given by $m\omega_\perp^2 r^2/2 $, where $\omega_\perp = 2\pi\cdot 100 \text{Hz}$. Dipoles are aligned by an external magnetic field along the $z$-direction. 
The phase diagram of dipolar gases is characterized by the ratio $\epsilon_{dd}$ between the dipolar and contact 2-body interaction strengths. Depending on its value, the system can be in a homogeneous superfluid, supersolid, or independent cluster phase, in which the superfluid long-range order along $x$ is lost (see, e.g.,~\cite{Roccuzzo1,F1,I1,S1}). Fixing the value of $\epsilon_{dd}$, we prepare the initial state of the system by evolving the extended GPE in imaginary time in the moving frame with velocity $V$. We change the reference frame by adding the term $-V P_x$ to the extended GPE Hamiltonian, where $P_x$ is the momentum operator along $x$. One can prove that such an approach unequivocally determines both the superfluid and the normal velocities \cite{SupplementaryMaterial}. In particular the velocity of the normal component coincides with the velocity of the moving frame ($v_n^0 = V$), while the appearance of the supercurrent is dictated by the energy minimization condition in the same moving frame.  The quantized superfluid velocity exhibits the first jump to a finite value $v_s^0=2\pi\hbar/(mL)$ at the critical value $V^{cr}=\vcr=\hbar\pi/(mL)=12.7\mu\text{m/s}$ (see also~\cite{tengstrand2021,Tengstrand2023} for the current generation in a ring).

It is worth noticing that both $v^0_n$ and $v^0_s$ could be in principle fixed in an independent way, by properly modifying the phase pattern $\phi$ with imprinting techniques. The value of the superfluid velocity is determined by the phase winding at the boundaries of the configuration, while the velocity of the normal component is determined by the space variations of the phase at a microscopic scale, fixed, in the case of a supersolid, by the inter-droplet distance (see \cite{SupplementaryMaterial}). However, such a state would not in general correspond to an equilibrium configuration in the moving frame.

\begin{figure}[!ht]
    \centering
    \includegraphics[width=\linewidth]{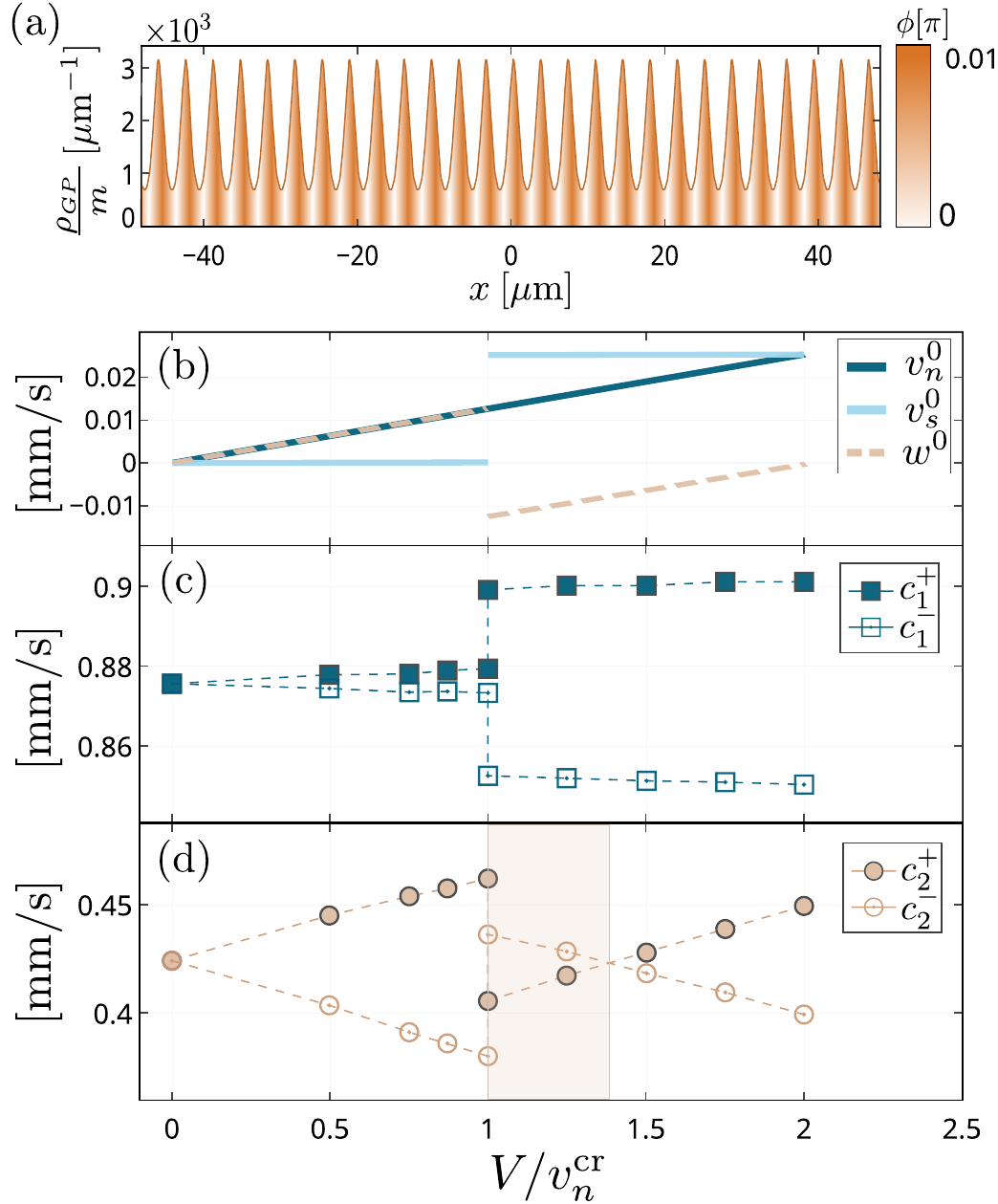}
    \caption{(a) 
    Transversely integrated density of a supersolid at $\epsilon_{dd}=1.398$  and its phase pattern at the tube's center (colorscale) moving with $v_n^0=\vcr$ and $v_s^0=0$. (b) Normal and superfluid velocities and their difference $w^0$ in a stationary state as a function of reference frame velocity $V$. Shapes of all three curves are independent of $\epsilon_{dd}$. Bottom panels present splitting of the (c) first and (d) second sound into four phonon modes for the system presented in (a). Within the shaded area, the Doppler shift of the second sound becomes negative. }
    \label{fig:sounds}
\end{figure}

In Fig.~\ref{fig:sounds}(a) we report the transversely integrated density $\rho_{GP}(x)$ and the phase  pattern $\phi(x,0,0)$ along the center of the tube for $v_n^0=\vcr$ and $v_s^0=0$. As mentioned above, one can see that the phase of a supersolid shows spatial modulations, whose gradients are related to the velocity of the normal component \footnote{For a physical interpretation and connection between the phase gradient and normal velocity in a moving supersolid we refer the Reader to the Supplemental Materials~\cite{SupplementaryMaterial}}. 
In Fig.~\ref{fig:sounds}(b) we show the values of $v_s^0$, with its characteristic jump, as a function of the reference frame velocity $V$ (for completeness also $v_n^0=V$ is shown)\cite{SupplementaryMaterial}. 

The supersolid phase at rest exhibits two  phononic modes, each of which splits into two in the presence of a current. As a consequence, as predicted by HD theory of supersolids (see discussion above), four different frequencies with linear dispersion $\omega = c|q|=c\cdot 2\pi/L$ are expected to occur. An example of Doppler splitting of the two sound modes $c_{1,2}$ in the supersolid phase is presented in Fig.~\ref{fig:sounds}(c-d). Both sounds experience an abrupt change of their speed at $V/\vcr=1$, i.e., when the phase of the order parameter starts hosting a quantum of circulation. Let us also notice that, remarkably, the Doppler shift for the sound speed $c_2$ can be negative, i.e. the velocity $c^+_2$ of the phonon copropagating with the current is smaller than $c^{0}_{2}$ (shaded region in Fig. \ref{fig:sounds}(d)). Formally, the negative Doppler shift occurs for $v_n^0/v_s^0 < 1-1/\delta_{2}$ \cite{SupplementaryMaterial}. 

In order to make a quantitative comparison  between the GPE and the predictions of  the linearized hydrodynamic equations,  we need to calculate the parameters entering the HD result Eq.~(\ref{eq:Doppler_ss}).  The parameters entering the speeds of sound in the absence of permanent currents ($f_s$, $\kappa$, $\beta$ and $\gamma$) have been discussed in previous works \cite{Sindik2023SoundSA,PlattSounds} (see also \cite{SupplementaryMaterial}). In particular, the strain-density coupling $\gamma$ has been shown to be very small for the supersolid phase of a dipolar gas~\cite{PlattSounds} and will be neglected in our analysis. The new parameters  $\rhon$ and $\rho'_s$ are determined numerically by marginally changing the total atom number $N$, while the derivative $\rhou$, quantifying the change in the normal density with respect to $\px u$, is calculated by varying the inter-droplet distance \cite{SupplementaryMaterial}.

\begin{figure}[!ht]
    \centering 
    \includegraphics[width=\linewidth]{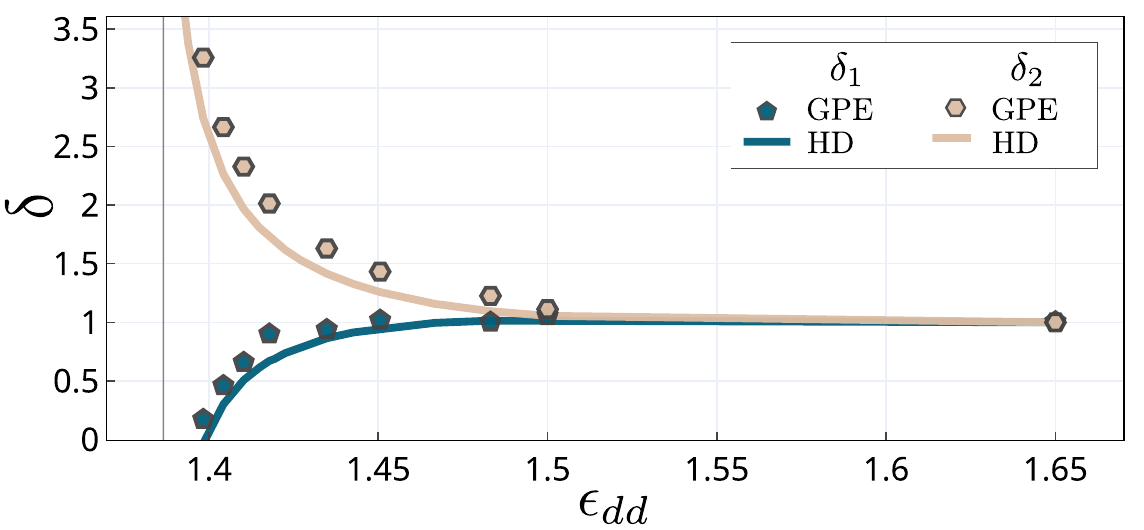}
    \caption{Anomalous Doppler shift of the first and second sound in the supersolid phase. Results of time-dependent simulations (markers) are compared with the hydrodynamic model (solid lines). The gray vertical line marks the superfluid-supersolid transition point.}
    \label{fig:d1d2}
\end{figure}

To test the hydrodynamic predictions, we extract the anomalous Doppler term   $\delta_{i}$ (see Eq.~(\ref{eq:Doppler_ss})) from the difference $\frac{1}{2}\left(c_{i}^+-c_{i}^-\right)$ calculated in the time-dependent GPE simulation. In Fig.~(\ref{fig:d1d2}) we show the comparisons between the HD theory (lines) and the numerical results (symbols). The good agreement between the hydrodynamic and time-dependent GPE predictions confirms the highly nontrivial form of the Doppler effect in supersolids. The effect is especially pronounced close to the superfluid-supersolid phase transition. 

\paragraph{Conclusions.} We have provided analytic and  numerical predictions for the zero-temperature Doppler effect  exhibited by ultracold atomic gases in the presence of density modulations caused by external periodic potentials, as well as by the spontaneous breaking of translational symmetry (supersolids). Analytic results are derived by linearizing the hydrodynamic theory of superfluids and supersolids, and compared with dynamic simulations based on the numerical solution of the GPE. We have shown that in single component superfluids anomalous Doppler shifts, following from the occurrence of a relative motion between the superfluid and normal velocities, can  take place even at zero temperature, thanks to the breaking of Galilean invariance, causing the appearance of the normal component. In the presence of an optical lattice, only fourth sound can propagate, and the Doppler effect takes a particularly simple form fixed by the density derivative of the superfluid density. For supersolids, where there exist sounds of hybrid  superfluid and crystalline nature, Doppler shifts exhibit highly non-trivial features, and for certain values of the parameters the shift can even have a negative sign.  We have considered the case of very elongated configurations which could be experimentally implemented confining the atomic gas in a ring geometry, and where one-dimensional hydrodynamic theory can be safely  applied.

\paragraph{Acknowledgement}
We acknowledge useful discussions with Jean Dalibard, and William D. Phillips. Stimulating discussions with Giovanni Modugno and the Pisa Dysprosium Lab are also acknowledged. This work has been supported by the Provincia Autonoma di Trento, Q@TN (the joint lab between University of Trento,
FBK-Fondazione Bruno Kessler, INFN-National Institute for Nuclear Physics and CNR-National Research
Council), CINECA consortium through the award under the ISCRA initiative,  for the availability of HPC resources. Part of the work was computed on ``Deeplearning cluster" supported by the initiative ``Dipartimenti di Eccellenza 2018-2022 (Legge 232/2016)" funded by the MUR.

\bibliography{bibliosupersolid}

\end{document}